# Impact of velocity and impact angle on football shot accuracy during fundamental trainings

Rahman Şahinler[1], Ömer Burak Göktaş[1], Berkay Mumcu[1], Damla Sen[2], Feyza Kocaturk[1], Huseyin Uvet[1*]

[1] Department of Mechatronics Engineering, Yildiz Technical University, Istanbul 34349, Turkey
[2] Uskudar American Academy, Istanbul 34664, Turkey
*Corresponding author email: huvet@yildiz.edu.tr

## Abstract

The purpose of this research is to create a machine learning-based smart coaching approach for football that can replace manual analysis with real-time feedback for trainers. In-depth analysis of football player data by humans is time-consuming, error-prone, and requires a lot of effort. This exploratory study demonstrates the feasibility of using a machine learning algorithm to enhance the effectiveness of player monitoring and training. The suggested approach uses machine learning to generate analytical insights and enable long-term monitoring of player performance. In the future, machine learning could use this technique to offer constructive criticism of football players. The system incorporates a homemade ball-throwing mechanism capable of launching the ball in a variety of directions and at varying velocities. The ball kicker is equipped with a gyroscope and accelerometer sensors for measuring velocity and acceleration. The gathered data is filtered initially, and then the data that has been processed is fed into the machine-learning algorithm. The algorithm will be trained on player performance data and will be able to provide real-time feedback to coaches on player performance and potential areas for improvement. Additionally, the system will be able to track player progress over time and provide coaches with a comprehensive view of player development. The ultimate goal is to improve player performance and reduce the workload for coaches by automating the analysis process.

## Introduction

The field of football has seen a growing interest in the application of machine learning in recent years. Studies have proposed the use of machine learning to enhance various aspects of the game, such as player performance, match predictions, injury predictions, strategy analysis, and game footage analysis.

A machine learning (ML) smart coaching strategy for football is an example of the type of research that has been conducted in this area [1-2]. Research in this area has proposed using machine learning to help with analysis and long-term player tracking. The proposed methods analyze data from a homemade ball-throwing mechanism fitted with gyroscope and accelerometer sensors to measure velocity and acceleration using machine learning. First, the collected data is filtered, and then the processed data is sent into the machine-learning algorithm. The results showed that this method might be used to improve player monitoring and training by giving coaches instantaneous feedback.

Another use of machine learning to sport is the analysis of football videos [3-4-5]. Research in this area has proposed applying deep learning methods to game footage in order to track players and analyze strategies for coaches. They relied heavily on convolutional neural networks (CNNs) to monitor player locations while also employing recurrent neural networks (RNNs) to decipher team strategies. These tests demonstrated that the suggested system achieved excellent levels of accuracy in player monitoring and tactical analysis, two areas that might guide coaches' decisions.

Football match prediction using machine learning is another excellent field [6-7-8-9]. These researches advocated for the use of machine learning to the task of predicting the outcomes of football matches by analyzing a variety of data, including player statistics and past game outcomes. The experiments used machine learning (ML) methods, such as a random forest algorithm, to forecast the outcomes of the matches, and the results demonstrated that the suggested system achieved high accuracy in match prediction. This can be used to aid fans and bettors in making informed predictions.

Studies have advocated using machine learning to anticipate injuries and evaluate the performance of football players by utilizing data from diverse sources, such as player medical records and physical performance measures [10-11-12]. Machine learning has been proven in these researches to be useful for assisting coaches and medical personnel in determining which players should be selected for a team and how they should be trained.

In addition, machine learning-based frameworks for football performance analysis proposed leveraging data from GPS-like tracking devices to analyze football player performance. The results of these research demonstrated that the proposed framework may be utilized to help coaches make decisions by extracting important information regarding player performance [13, 14]. The study's authors proposed a machine learning-based approach to analyze football game film for information relevant to coaches, such as player tracking and tactical analysis [15, 16]. The research tracked the locations of the players on the field and analyzed the teams' strategies using a combination of computer vision and machine learning. High levels of accuracy in player monitoring and tactical analysis were found using the suggested approach, which can be utilized to aid coaches in their decision making [17, 18].

These studies demonstrate that machine learning may be applied to improve several facets of football, including player performance, match prediction, injury prediction, strategy analysis, and analysis of game footage. These studies provide preliminary evidence that machine learning may prove useful in future multi-perspective analyses of the football sport. Machine learning's pattern-recognition capabilities and massive data processing power make it a useful tool for coaches looking to boost their teams' overall performance by individually boosting the stats of their most valuable players. In addition, smart coaching solutions powered by machine learning can give instantaneous feedback to coaches, cutting down on time-consuming manual analysis while also boosting productivity. Coaching staffs can gain a more complete picture of their players' growth through the use of machine learning-based technologies that allow for continuous monitoring of player statistics throughout time.

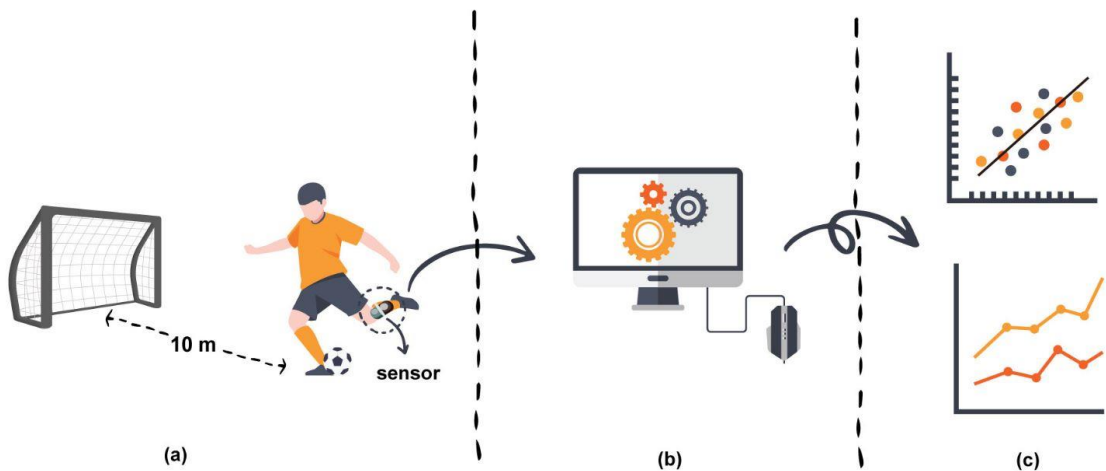

*Figure 1, (a) data collection using inertial sensors (b) data processing of collected data (c) data visualization and shot classification*

Our goal is to create a football teaching system for training purposes that can contribute novel and cost-effective solutions to this emerging industry. This study indicates using inertial measurement sensors in tandem with machine learning to assess a fundamental football kicking motion by measuring outputs like ball meeting point, ball velocity, and strike rate (Figure 1). The goal of our research is to create an AI-based DSS that can analyze football kicks and provide feedback on the player's play.

## Concept Study and Results

### Data Collection

Data collection protocol has been standardized and applied to all experimental phases; those are
- The distance between the football player and the goal set to 10 meters (Figure 2-a).
- The location where the sensor is attached, is under the calf over the ankle (Figure 2-b).
- The same ball is used in each measurement.
- Shooting times are the same for each measurement.
- Polar H10 is attached to chest synchronized with gyroscope and accelerometers.

IMU electrical circuit and battery are placed on the player's feet as shown in the figure 1. The player has completed standardized shooting stages with 7 seconds per measurement. In each measurement, a data flow amount of approximately 7Hz could be reached. These results were repeated 1000 times on 5 different people and the results were recorded.

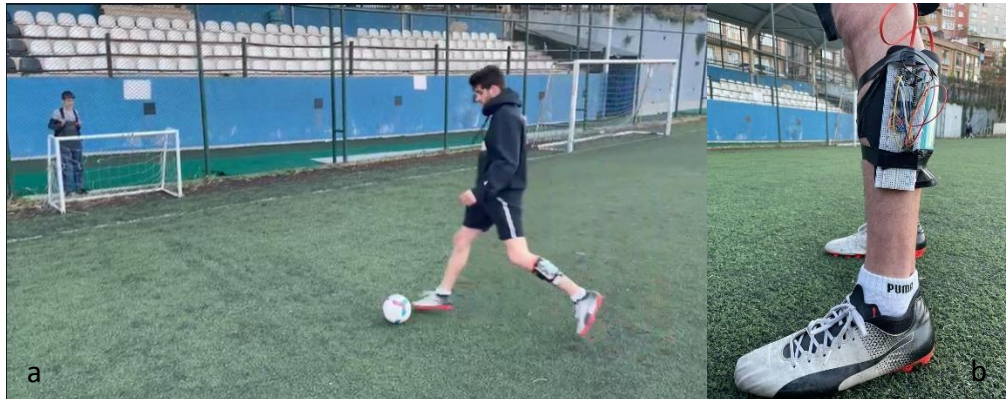

*Figure 2, (a) - Data collection in the field, (b) sensor positioning*

### Data Analysis

In our study, we used linear regression to model the connection between a player's 3-axis acceleration data (the independent variable) and the outcome of a shooting challenge (the dependent variable) (i.e., the independent variables). First, we measured the 3-axis data with sensors worn by the players and used an accelerometer and gyroscope to calculate the precision of the shots. Following this, we compiled all of the data into a dataset for use in training a linear regression model. Once the linear regression model was trained, it was used to new sensor data to predict the results of the shooting. Five different players' sensor data was collected and fed into a linear regression model to determine each one's chances of success in the shooting challenge. We refer to this optimal shooting position as "ground truth," and the computer finds the best-fitting line across the data points to get there.

However, it is important to note that there were challenges with data collection at every shoot, including (1) loss of data received via Bluetooth, (2) attachment failure of maintaining the position of the electronic circuit on player leg, (3) data filtering during sensor acquisitions, (4) player positioning to the ball.

### Experimental Results

Data has been processed step by step to establish a meaningful correlation between the player's shooting accuracy and the data collected using an IMU sensor. The graphs in figure 3 of the gyroscope and acceleration data obtained from the IMU sensor are given in the x, y, and z axes. These data are raw and have not undergone any filtering process. A randomly selected one of the shots taken as a sample of data is selected.

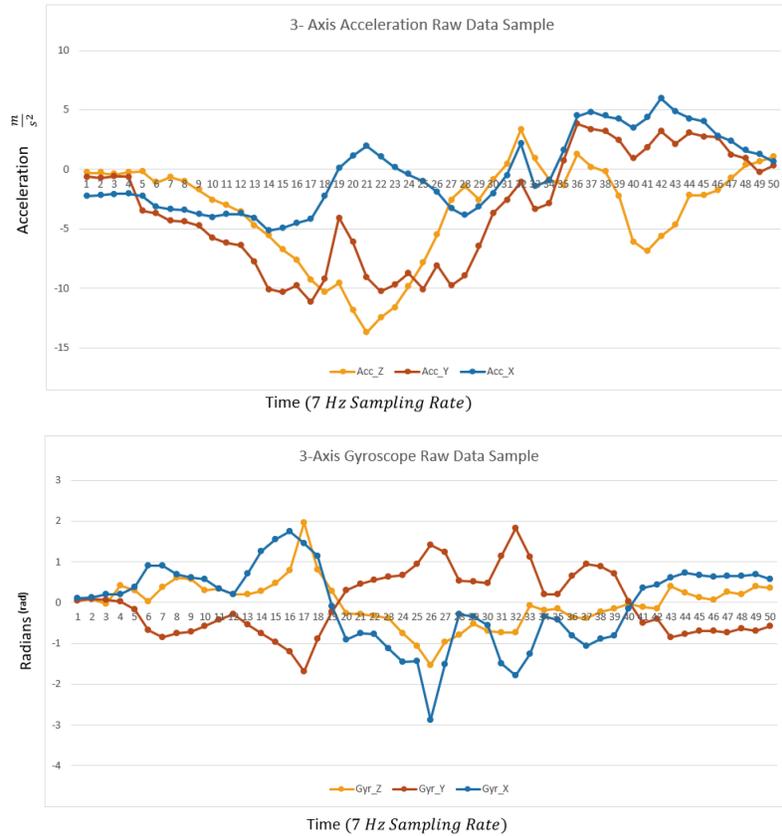

*Figure 3, Gyroscope and acceleration data obtained from the IMU sensor*

The gyroscope data obtained must be preprocessed because it contains high level of noise. Based on the results of literature research performed before using the gyroscope data, it has been filtered using the Complementary Filter Algorithm. The choice of this filtering method and the data flow rate (7 Hz) have been determined. A dataset consisting of approximately 30,000 data points was input into a linear regression algorithm for each shot. After passing through this algorithm, the goal is to find the optimal shape of the graphs obtained during the shot and use them as a ground truth model for comparative shot analysis. Figure 4 shows the Acceleration and Gyroscope data used to create the "Ground Truth" model on three axes.

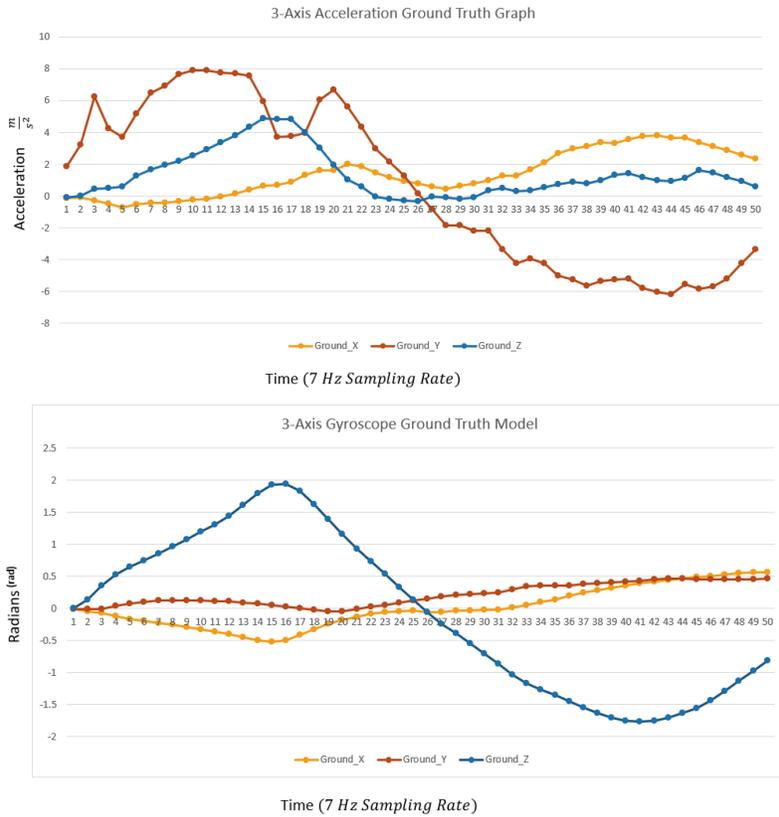

Figure 4, Ground Truth" model on three axes

The Y-axis in the Acceleration data, which represents the shooting direction, and the Z-axis in the Gyroscope data, which indicates the deviation of the player's leg during the shot, are used as the basis for shot analysis. The Acceleration data (Figure 5) reveals the probability of a successful shot based on the deviation ratio from the Ground Truth model, which is the optimum shot hardness, during the shot process when the player hits the ball along the shooting axis. The Gyroscope data also measures the deviation of the shot axis from the X-axis with the angular deviation of the deviation during the shot process. The similarity ratio of the results after the shot to the Ground Truth Model directly affects the success rate of the shot. The graphs used for comparison are given Figure 6.

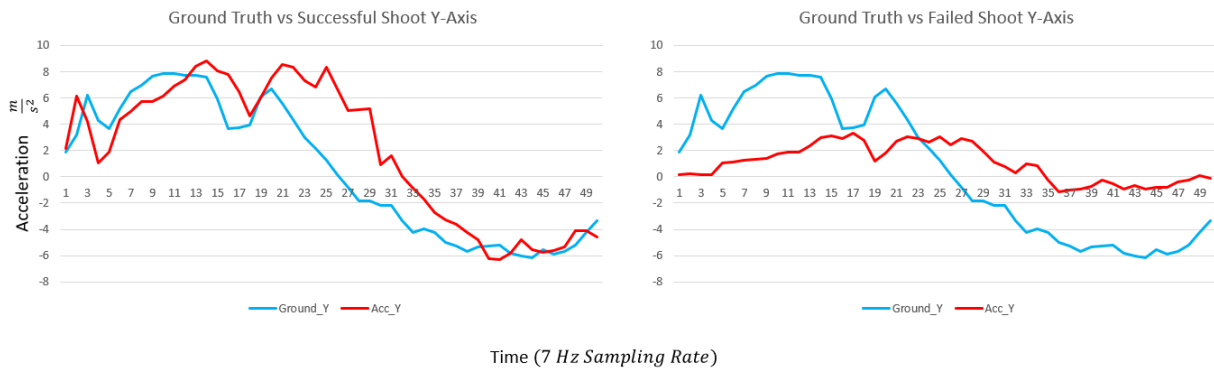

Figure 5, Successful attempts (Acc_Y) generate results that are relatively near to the ground truth line (Gound_Y), as shown in the graphs (Left) while unsuccessful shots exhibit a distinct curve (right).

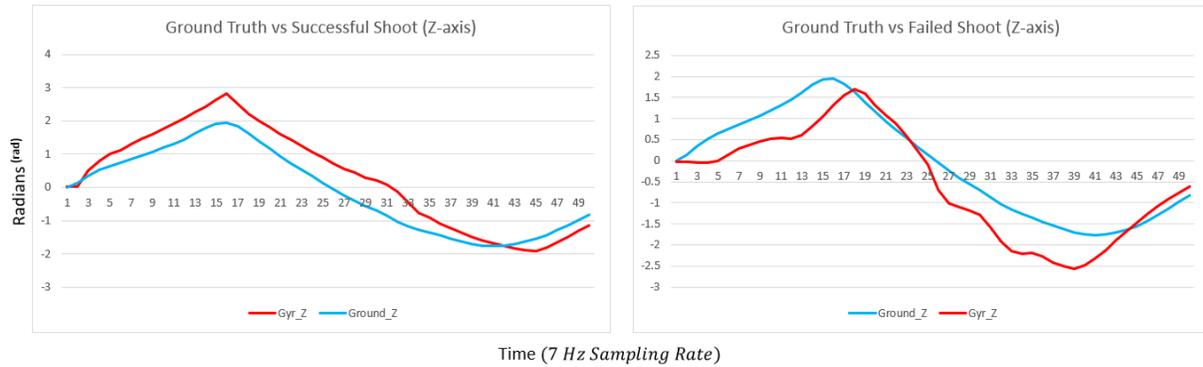

*Figure 6, Data from the gyroscope and its correlation to a successful shot*

Figure 7 shows a similar phenomenon, with the Gyroscope data correlating with the Ground Truth model for accurate shots but not for inaccurate ones. The entire graph used for the shot contains the total data amount of 7 seconds during the shot process. This amount is quite excessive for a meaningful comparison. The necessary part for comparison is indicated as the shooting phase. The parallelism ratio of the graphs during this process indicates the success or failure of the shot.

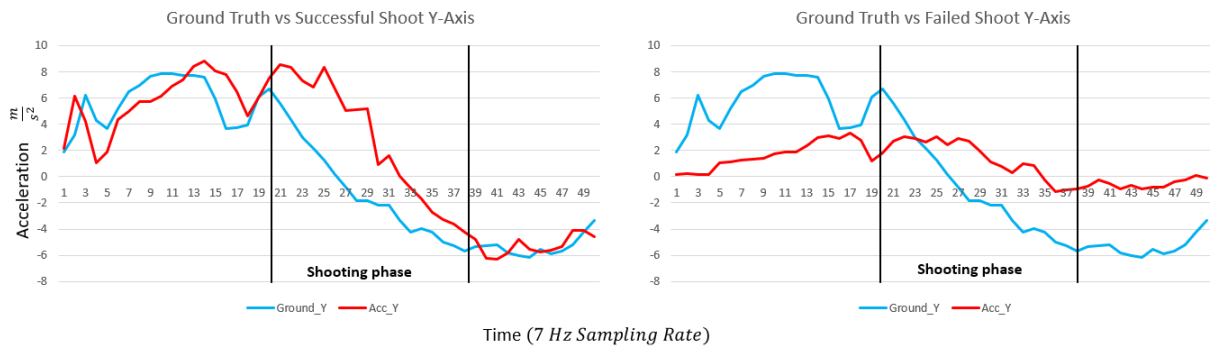

*Figure 7, Shooting phase is the moment when player hit the ball. Left graph indicates successful shoot while the right one indicates failed one.*

The highlighted region represents a comparison of the successful and unsuccessful shots' acceleration data with the Ground Truth model during the designated shooting phase (Figure 8). Successful attempts exhibit close proximity within the allocated time, while complex regions characterize the accelerated over time attempts that are more likely to fail.

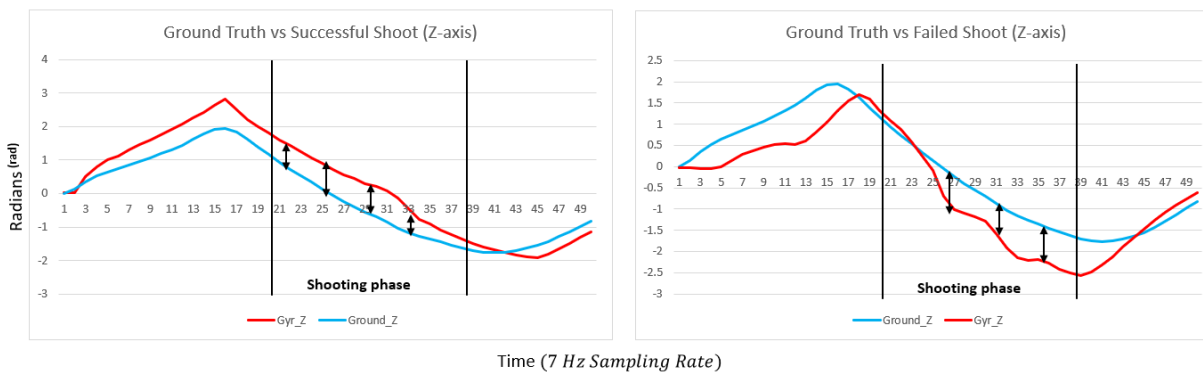

*Figure 8, When there is a wider gap between the blue and red waves, the shot has a reduced chance of success.*

Each chart reveals the different characteristics when the shooting phase is analyzed. Gyroscope readings are closer to the ground truth when the shot is successful, and they deviate more when the shot misses.

Discussion and Conclusion

The research was conducted by analyzing shot data from a variety of players with varying attributes utilizing a standard shot idea. Each participant took a series of shots according to a predetermined shot concept, and statistics were recorded as they occurred. An approximate 30.000 data collection was cleaned up to include all relevant characteristics and made ready for analysis. Through the use of a Linear Regression Algorithm, the statistics of the best shots were extracted, allowing for the development of a Ground Truth model. Analyses of both the shot data and the Ground Truth model were performed. The critical dimensions of the shot and the time frame that characterizes the shooting procedure were employed in this comparison.

The study contrasted the resulting Ground Truth model with the velocity and gyro data from the randomly taken shot. An increase in both shot success probability and shot resemblance to the Ground Truth model was noted. If the data for the relevant axis' velocities deviates from the Ground Truth model, it means that the shot process did not attain sufficient force, which increases the shot's likelihood of failing. It was found that high angular deviation during the shooting process was correlated with a gyro reading that deviated from the Ground Truth model.